\documentclass[showpacs,showkeys]{revtex4}%
\usepackage[colorlinks=true]{hyperref}

\AtBeginDocument{
	\hypersetup{
		urlcolor=cyan,
		citecolor=magenta,
		linkcolor=orange,
		anchorcolor=green,
	}%
}
\usepackage{amssymb}
\usepackage{amsmath}
\usepackage{amsfonts}
\usepackage{graphicx}
\usepackage{ulem} 

\usepackage{cancel}
\usepackage[usenames]{xcolor}
\usepackage{epstopdf}
\usepackage{extarrows}
\providecommand{\rbr}[1]{\left( #1 \right)}%
\providecommand{\sqbr}[1]{\left[ #1 \right]} %

\begin{document}

\title[ ]{R\'enyi entropy yields artificial biases not in the data and incorrect updating due to the finite-size data}
\author{$^{1}$Thomas Oikonomou}
\email{thomas.oikonomou@nu.edu.kz}
\author{$^{2}$G. Baris Bagci}
\affiliation{$^{1}$Department of Physics, School of Science and
Technology, Nazarbayev University, Astana
010000, Kazakhstan}
\affiliation{$^{2}$Department of Physics, Mersin University, 33110 Mersin, Turkey}


\begin{abstract}
We show that the R\'enyi entropy implies artificial biases not warranted by the data and incorrect updating information due to the finite-size of the data despite being additive. It is demonstrated that this is so because it does not conform to the system and subset independence axioms of Shore and Johnson. We finally show that the escort averaged constraints do not remedy the situation.       
\end{abstract}

\eid{ }
\date{\today }
\startpage{1}
\endpage{1}
\maketitle

Boltzmann-Gibbs-Shannon (BGS) entropy and its quantum mechanical counterpart von Neumann entropy has found applications in many diverse fields of science \cite{S1,S2,S3,S4,S5,Q1,Q2,Q3,Q4,Q5,Q6,N1,N2,N3,W1,W2}. Its use is rigorously rationalized through e.g. the coding theorems \cite{code} or the Khincin axioms \cite{K}. On the other hand, the entropic functional does not suffice for many practical issues and one also needs the concomitant equilibrium probability distribution. This distribution is indeed known in statistical mechanics for closed systems as the canonical one. However, it was Jaynes who obtained the same distribution through a procedure named entropy maximization procedure (MaxEnt) by relying on the information theory \cite{Jaynes1}. From then on, obtaining the equilibrium entropy became an inference problem in statistical estimation theory.

Whether this inference is consistent or not is warranted by the Shore-Johnson (SJ) criteria \cite{SJ}. These criteria consist of four axioms which ensure choosing the suitable entropic functional if one aims self-consistent inferences from the data. It is only afterwards that this functional can be safely maximized. In this sense, SJ axioms are related to the pre-maximization scheme. If a functional violates one of the SJ axioms, then it cannot be securely used for the MaxEnt procedure \cite{Presse5}.

On the other hand, a multitude of entropies is recently introduced and applied in diverse fields of research \cite{entropy1,entropy2,entropy3,entropy4,entropy5,Bagci1}. The main and common claim of these entropies is that they are suitable for the cases where the number of accessible states does not grow exponentially \cite{Canosa,Taruya,Bagci2,Bagci3}. One of the mostly used entropy measures is the R\'enyi measure. It is used extensively for quantum information, generalized statistical mechanics and quantum gravity. Then, this begs the question whether the R\'enyi entropies conform to some or all of the axioms, thereby allowing consistent inferences.

There are four axioms forming the SJ criteria \cite{SJ,Presse,Presse2,Presse3,Presse4}. The first one is the axiom of uniqueness whose requirement is the concavity (convexity) of the (relative) entropy measure to warrant a unique maximum. The second one is called the invariance axiom and requires the inferences to be independent of the adopted coordinate system. The third axiom is the subset independence. It states that treating an independent subset of system states in terms of separate conditional probabilities and in terms of joint probabilities are equivalent. From a different point of view, it implies that the MaxEnt procedure should yield the same results whether the number of states is fully accounted for or not. The last axiom i.e. system independence warrants the absence of biases when the systems are independent if the data do not provide evidence otherwise. In particular, for systems $A$ and $B$ with probabilities $\left\lbrace u_{i}\right\rbrace $ and $\left\lbrace v_{j}\right\rbrace $, respectively, their combination creates new bins with the joint probability $p_{ij} = u_{i}v_{j}$.

It is worth emphasis that SJ axioms have precedence to the MaxEnt, since it is interested in consistent and unbiased inference i.e. pre-maximization \cite{Presse3}. These axioms are about deriving a functional which is only later to be maximized. Skipping this vital step implies the failure of both MaxEnt and post-maximization related arguments such as extensivity etc. but not vice versa.         

It has recently been shown that the nonadditive $q$-entropies only violate the probability independence axiom \cite{Presse}. It might be argued that the same result should also hold for the R\'enyi entropy, since the nonadditive $q$- and additive R\'enyi entropies are monotonic functions of each other and therefore should result in the same probability distributions after the maximization procedure. However, since the SJ axioms are related to the pre-maximization scheme and choose the functional which is later to be maximized, one must consider the R\'enyi entropy as a different and important case study compared to the nonadditive $q$-entropies.

In order to check the SJ axioms for the R\'enyi entropy, we begin with the following functional 
\begin{eqnarray}\label{departure}
H_{q} \left( \left\lbrace p \right\rbrace  \right) - \lambda \rbr{\sum_{k} p_{k} a_k-\overline{a}}
\end{eqnarray}  
where $\lambda$ and  $\overline{a}$ are Lagrange multiplier and the measured average of the quantity $a$, respectively. The normalization constraint is omitted for simplicity. The term $H_{q}$ is the R\'enyi entropy given as
\begin{eqnarray}\label{renyi}
H_{q}(\{p\})=\frac{1}{1-q}\ln\rbr{\sum_{i=1}^{n} p_{i}^q}\,,
\end{eqnarray} 
where $q$ is the generalization parameter.

The first axiom can be easily checked, since the R\'enyi entropy $H_{q}$ is concave for $q\in(0,1)$. Therefore, the uniqueness axiom is satisfied by the R\'enyi entropy in the interval $q\in(0,1)$.

The second axiom i.e. the invariance axiom requires some clarification. The discrete R\'enyi entropy (or discrete BGS entropy for that matter) is not coordinate invariant. Therefore, one generalizes them to the continuous case in the form of relative entropy so that one safely optimizes the associated relative entropy. This relative entropy expression, if it exists, can be seen to be coordinate invariant. For the BGS entropy, this is given by the Kullback-Leibler relative entropy, for example. Therefore, the crux of this axiom is to obtain the continuous version of the discrete entropy measure so that it will be coordinate invariant (see Ref. \cite{Bagci5} for details). To this aim, consider the discrete R\'enyi entropy as the point of departure and following the Jaynes' approach, substitute $p_i=\rho(x_i)\Delta x_i=\rho(x_i)/[nm(x_i)]$ into the discrete expression to obtain
\begin{eqnarray}
\nonumber
\lim_{n\to\infty}H_{q}(\{p\}) &=& \lim_{n\to\infty}\frac{1}{1-q}\ln\rbr{\sum_{i=1}^{n}\sqbr{\frac{\rho(x_i)}{m(x_i)}}^q m(x_i) \Delta x_i} +\lim_{n\to\infty}\ln(n)
\end{eqnarray}
so that the continuous R\'enyi (relative) entropy $H_{q}[\rho||m]$ reads 
\begin{eqnarray}
\nonumber
H_{q}[\rho||m] = \frac{1}{1-q}\ln\rbr{\int_a^b\sqbr{\frac{\rho(x)}{m(x)}}^q m(x) \mathrm{d}x}\,,
\end{eqnarray}
where the additive term is omitted exactly as in the BGS case, since the entropy change is what matters. Note that $m(x)$ is now a measure on the continuous version of the state space.

The third axiom in the SJ set of criteria is the subset independence. For the R\'enyi entropy to conform to this axiom, the following expression should be equal to zero \cite{Presse5}
\begin{eqnarray}\label{SubSetInd}
\frac{\partial}{\partial p_\ell}\rbr{\frac{\partial}{\partial p_{k}} - \frac{\partial}{\partial p_{j}}}\rbr{H_{q}-\lambda_a \sum_{i=1}^{n}p_i a_i}.
\end{eqnarray}
Substituting the R\'enyi entropy $H_{q}$, the expression above reads 
\begin{eqnarray}
 \frac{\partial}{\partial p_\ell} \sqbr{\frac{q(p_k^{q-1}-p_j^{q-1})}{(1-q)\sum_{i=1}^{n}p_i^q}-\lambda_a a_k +\lambda_a a_j}
\end{eqnarray}
which, after a little algebra, yields
\begin{eqnarray}
\frac{q(p_k^{q-1}-p_j^{q-1})}{1-q}\frac{\partial}{\partial p_\ell}\rbr{\sum_{i=1}^{n}p_i^q}^{-1}.
\end{eqnarray}
Since this expression is not zero in general, we conclude that the R\'enyi entropy violates the third SJ axiom, i.e., the subset independence axiom. The implication of this violation is that the inference drawn from the updating procedure strongly depends on the set $S\subset D$ of limited data currently available to us, where $D$ is the maximum state set of the system under scrutiny.

To check the last axiom, namely, the system independence axiom, we consider the following functional
\begin{eqnarray}\label{SJR_lin01}
\Lambda=\frac{1}{1-q}\ln\rbr{\sum_{i,j}f(p_{ij})} - \lambda_a \rbr{\sum_{i,j}p_{ij} a_i-\overline{a}} - \lambda_b \rbr{\sum_{i,j}p_{ij} b_j-\overline{b}}\,.
\end{eqnarray}
The maximization condition $\delta \Lambda=0$ yields
\begin{eqnarray}
\frac{f'(p_{ij})}{(1-q)\sum_{k,l}f(p_{kl})}-K_{ij}=0 \,,
\end{eqnarray}
where $K_{ij}\equiv \lambda_a a_{i}-\lambda_b b_{j}$. Summing over all indices, we obtain the relation
\begin{eqnarray} 
K f'(p_{ij})=K_{ij}\sum_{k,l}f'(p_{kl})
\end{eqnarray}
with $K\equiv \sum_{i,j} K_{ij}$. Taking first the partial derivative $\frac{\partial}{\partial u_i}$ of the equation above
\begin{eqnarray}
K f''(p_{ij})\frac{\partial p_{ij}}{\partial u_i}=K_{ij} \sum_{l} f''(p_{il})\frac{\partial p_{il}}{\partial u_i}
\end{eqnarray}
and then  the partial derivative $\frac{\partial}{\partial v_j}$, we obtain
\begin{eqnarray}
(K-K_{ij}) \sqbr{f'''(p_{ij})\frac{\partial p_{ij}}{\partial u_i}\frac{\partial p_{ij}}{\partial v_j} + f''(p_{ij}) \frac{\partial^2 p_{ij}}{\partial u_i\partial v_j}}=0\,.
\end{eqnarray}
Since the first term in the parentheses is non-zero, the second one in the brackets must be equal to zero. Using the multiplicative joint probability $p_{ij}=u_iv_j$, the expression within the brackets yields 
\begin{eqnarray}
p_{ij}\,f'''(p_{ij}) + f''(p_{ij}) =0\,.
\end{eqnarray}
This differential equation is known to yield the BGS entropy (see Eq. (5) in Ref. \cite{Presse}). Therefore, we conclude that R\'enyi entropy also violates the system independence axiom.

A crucial point is that the R\'enyi entropy is often maximized through the so-called escort distributions to obtain the averaged quantities. These distributions are of the form $\frac{\sum_{i=1}^{n}p_i^qa_i}{\sum_{k=1}^{n}p_k^q}$. As we can show, the use of the escort distribution can not provide conformity of the R\'enyi entropy to the SJ axioms neither. The first two axioms i.e. the uniqueness and coordinate invariance are still satisfied by the R\'enyi entropy, since the changes in the averaging scheme leave the entropy expression unchanged. However, the subset independence axiom relies on how we average the constraints. In fact, with the escort averaging, Eq. (\ref{SubSetInd}) becomes
\begin{eqnarray}
\frac{\partial}{\partial p_\ell}\rbr{\frac{\partial}{\partial p_j} - \frac{\partial}{\partial p_k}}\sqbr{H_{q} - \lambda \frac{\sum_i p_i^q a_i}{\sum_i p_i^q}}\,.
\end{eqnarray}
In order to check whether the subset independence axiom is satisfied, the above relation should be equal to zero. After a little algebra, it reads

\begin{eqnarray}
(p_j^{q-1}-p_k^{q-1}) \left[\frac{1}{q-1}+\lambda \rbr{a_\ell -2\frac{\sum_i p_i^a a_i}{\sum_i p_i^q}} \right] +\lambda  (p_j^{q-1}a_j-p_k^{q-1}a_k)
\end{eqnarray}
which is zero only for $q \rightarrow 1$ i.e. in the BGS entropy. As a result, we conclude that the escort averaged constraints do not prevent the R\'enyi entropy from violating the subset independence axiom. 

The next and last axiom to be checked is the system independence axiom. Similar to Eq. (\ref{SJR_lin01}) but now with escort averages, we write 
\begin{eqnarray}
\Lambda=\frac{1}{1-q}\ln\rbr{\sum_{i,j}f(p_{ij})}-\lambda_a\sqbr{ \frac{\sum_{i,j}(p_{ij})^qa_i}{\sum_{k,l}(p_{kl})^q}-\overline{a}}-\lambda_b\sqbr{ \frac{\sum_{i,j}(p_{ij})^qb_j}{\sum_{k,l}(p_{kl})^q}-\overline{b}}
\end{eqnarray}
Through $\delta \Lambda=0$, we have
\begin{eqnarray}
0=\frac{q}{1-q} \frac{f'(p_{ij})}{\sum_{k,l}f(p_{kl})}-\lambda_a q\frac{(p_{ij})^{q-1}\left(a_i-\frac{\sum_{i,j}(p_{ij})^qa_i}{\sum_{k,l}(p_{kl})^q}\right)}{\sum_{k,l}(p_{kl})^q}-\lambda_b q\frac{(p_{ij})^{q-1}\left( b_j - \frac{\sum_{i,j}(p_{ij})^qb_j}{\sum_{k,l}(p_{kl})^q} \right)}{\sum_{k,l}(p_{kl})^q}.
\end{eqnarray}
Multiplying this equation with $p_{ij}$ and summing over $i,j$'s we see
\begin{eqnarray}
0=\sum_{i,j}p_{ij}f'(p_{ij})
\end{eqnarray}
where again $q\in(0,1)$. Applying now the derivative $\frac{\partial}{\partial u_i}$
\begin{eqnarray}
0&=&\sum_{k,l}\sqbr{f'(p_{kl}) + p_{kl}f''(p_{kl}) }\frac{\partial p_{kl}}{\partial u_i} 
=\sum_{l}\sqbr{f'(p_{il}) + p_{il}f''(p_{il}) }\frac{\partial p_{il}}{\partial u_i}
\end{eqnarray}
and then the derivative $\frac{\partial}{\partial v_j}$ we have
\begin{eqnarray}
0 &=& \sqbr{2f''(p_{ij}) + p_{ij}f'''(p_{ij}) }\frac{\partial p_{ij}}{\partial v_j} \frac{\partial p_{ij}}{\partial u_i} + \sqbr{f'(p_{ij}) + p_{ij}f''(p_{ij}) }\frac{\partial^2 p_{ij}}{\partial v_j\partial u_i}\,.
\end{eqnarray}
Substituting the probability multiplication $p_{ij}=u_iv_j$ we finally have
\begin{eqnarray}
0&=&f'(p_{ij}) + 3p_{ij}f''(p_{ij}) + p_{ij}^2f'''(p_{ij})
\end{eqnarray}
whose solution reads
\begin{eqnarray}
f(p_{ij})=c_1+c_2 \ln(p_{ij})+\frac{c_3}{2}\ln^2(p_{ij})\,.
\end{eqnarray}
Apparently, the R\'enyi entropy violates the probability independence axiom, since the measure $\frac{1}{1-q}\ln\rbr{\sum_{i,j}f(p_{ij})}$ with the above expression i.e., $f(p_{ij})=c_1+c_2 \ln(p_{ij})+\frac{c_3}{2}\ln^2(p_{ij})$, is not the R\'enyi entropy.

Before concluding, one can also consider a related entropy expression i.e. the homogeneous entropy \cite{Arimoto,Boon} in terms of the SJ axioms
\begin{eqnarray}
S^H_q=\frac{\left[\sum_i p_i^{1/q}\right]^q -1}{q-1}.
\end{eqnarray}
The homogeneous entropy can easily be checked to satisfy the first axiom due to its concavity. However it fails the second one: retracing the aforementioned steps regarding the second SJ axiom, we obtain
\begin{eqnarray}
S^H_q = \frac{1}{q-1} \Bigg\{n^{q-1}\left[\sum_i m(x_i) \left(\frac{\rho(x_i)}{m(x_i)}\right)^{1/q}\Delta x_i  \right]^q-1\Bigg\}
\end{eqnarray}
As can be seen,  the expression does not converge as $n\to\infty$. In other words, the homogeneous entropy does not have a continuous expression and is bound to be used only for the discrete cases. Regarding the third axiom i.e. subset independence, we find
\begin{eqnarray}
\frac{\partial}{\partial p_\ell}\left(\frac{\partial}{\partial p_k} - \frac{\partial}{\partial p_j}\right)\left(S_q^H-\lambda_a \sum_i p_i a_i\right)
&=& \frac{q\left(p_k^{\frac{1}{q}-1} - p_j^{\frac{1}{q}-1}\right)}{q-1} \left(\sum_i p_i^{1/q}\right)^{q-1} \frac{\partial}{\partial p_\ell} \left(\sum_i p_i^{1/q}\right)
\end{eqnarray}
which is generally nonzero apart from the well-known limit $q\to1$. Thus, the subset independence  axiom is also violated by the homogeneous entropy. The last axiom i.e. system independence yields the relation $(1/q-1)p_{ij}=0$ when one follows the above procedure. Apparently, this relation is satisfied only for $q=1$, therefore finally showing that the homogeneous entropy only satisfies the first i.e. uniqueness axiom of the SJ criteria, violating the other three axioms explicitly. Following similar steps, one can see that $S_q^H$ does not conform to the SJ axioms even with  the escort averaging. 

To sum up, SJ criteria should be satisfied by any entropy measure if it is to be used for consistent inferences from the data. The R\'enyi entropy violates both system and subset independence axioms, thereby resulting in the artificial biases not present in the data and updating information dictated by finite-size effects, respectively. As a result, it cannot be used to draw inferences in a consistent manner. On the other hand, the nonadditive $q$-entropies violate only the system independence axiom \cite{Presse}. This might be surprising at first, since these two entropies are monotonic functions of one another. After all, they yield the same type of probability distributions after the entropy maximization procedure is carried out. The resolution of this apparent conflict is at the heart of the SJ axioms, since these axioms, even before the maximization procedure, determine whether the functional under scrutiny is worth maximizing at all. Despite the violation of the SJ criteria, one might nevertheless choose to maximize an entropy measure and obtain the concomitant optimum distribution. However, any such violation is bound to cast doubt on the practical use of the entropy expression such as yielding e.g. artificial biases or inapplicability for continuous physical systems even before one hopes for a consistent thermodynamics \cite{Bagci4,Plastino}. Since SJ criteria are related to the pre-maximization scheme \cite{Presse3}, the R\'enyi entropy is an important case by itself whose features cannot be deduced from those of the nonadditive $q$-entropies. Note that the R\'enyi entropy is sometimes used together with a different averaging scheme i.e. so called escort averaging. However, this change in averaging constraints neither prevents the R\'enyi entropy from violating these two important axioms. Finally, one might try to extend the scope of one of the axioms, for example, the probability independence axiom \cite{Jizba}. However, this does not suffice, since the R\'enyi entropy would still violate the subset independence axiom. In short, the R\'enyi entropy can be used for various purposes but is inadequate for consistent inferences to be drawn. Having considered also the homogeneous entropy and shown that it also fails the SJ criteria, we emphasize that the SJ criteria single out the Shannon entropy as the unique measure to draw consistent inferences if one adopts the linear averaging scheme.

\begin{acknowledgments}
	The authors acknowledge the ORAU grant entitled ``Casimir light as a probe of vacuum fluctuation simplification" with PN 17098. T.O. acknowledges the state-targeted program ``Center of Excellence for Fundamental and Applied Physics" (BR05236454) by the Ministry of Education and Science of the Republic of Kazakhstan. G.B.B. acknowledges support from Mersin University under the project 2018-3-AP5-3093. 
\end{acknowledgments}


\end{document}